\documentclass[12pt]{article}
\usepackage{a4wide}

\usepackage{pst-all}
\input epsf

\begin{document}
\author{Bernard Linet \thanks{E-mail: linet@lmpt.univ-tours.fr} \\
\small Laboratoire de Math\'ematiques et Physique Th\'eorique \\
\small CNRS/UMR 6083, F\'ed\'eration Denis Poisson \\
\small Universit\'e Fran\c{c}ois Rabelais, 37200 TOURS, France}

\title{\bf The south-pointing chariot on a surface}
\date{}
\maketitle

\thispagestyle{empty}

\begin{abstract}

We examine closely the motion of the south-pointing chariot on a surface by taking into account the fact that 
both wheels have to roll without slipping. We firstly develop a procedure of successive approximations. 
In the limit where the distance between the wheels
tends to zero, we find again the result due to Santander that the pointer is parallel transported. This is
longer true generally within the first-order approximation. We secondly determine in an exact manner the motion
of the south-pointing
chariot on the sphere. Then, we define a pointer and we prove that this one is parallel transported exactly.

\end{abstract}

\section{Introduction}

Historically, the Chinese south-pointing chariot is a two-wheeled chariot which is surmounted by a human 
figure whose the arm
serves as a pointer. At least as it is recorded in ancient Chinese texts, the pointer has the property to
always show the same direction when the south-pointing chariot is moving. What is the mechanism, or the
different mechanisms, invented by the Chinese in such a way that the pointer keeps the same direction?
It is not sure that the answer can be given by the sinologists at present \cite{jing}.

Facing difficulties of a historical reconstruction, a possible approach is to ask what mechanisms
allow a south-pointing chariot. This is the position of George Lanchester, an British engineer in the
automobile industry. In a conference given in 1947 \cite{lanc}, he argues that this mechanical device should be a
planetary gear differential, as in a car, without actually referring to original texts. At present, there is
a classification of possible devices \cite{yan}.

We briefly recall how the differential gear of a south-pointing chariot runs in the plane by
considering a simple example. The chariot has two wheels of radius $R$ separated by a distance $D$.
A part of the motion is represented by the paths
of the two wheels which are respectively the two arcs of circle of radius $a$ and $a-D$ for an 
angle $\delta \psi$ going anticlockwise. In our example, but it is not essential, we assume that
$D\leq a$ and in the case $D=a$ the left wheel revolves around the center of the circle. For a rolling
without slipping, the angles of the rotating motion of the right and left wheels are respectively
$$
\delta \theta_R=\frac{a}{R}\delta \psi \quad {\rm and} \quad \delta \theta_L=\frac{a-D}{R}\delta \psi .
$$ 
A differential gear combines the data $\delta \theta_R$ and
$\delta \theta_L$ to give an angle $\delta \phi$ as follows
\begin{equation}\label{diff1}
\delta \phi =\frac{1}{2}\left( \delta \theta_R -\delta \theta_L \right) . 
\end{equation}
Taking into account $\delta \theta_R$ et $\delta \theta_L$, we obtain
\begin{equation}\label{diff2}
\delta \phi =\frac{D}{2R}\delta \psi .
\end{equation}
We denote $\phi$ the angle between the pointer and the axle of the wheels.
After this motion, if the new angle of the pointer is $\phi +\delta \psi$ then this one has
conserved the same direction. Therefore, we adopt the condition $D=2R$ in relation (\ref{diff2}) and the role of
the differential gear is to add $\delta \phi =\delta \psi$ to the angle $\phi$ continually.

In fact, this mechanical device is not efficient over a long distance because a weak difference
between the radius of the wheels induces cumulative errors \cite{coal}. Nevertheless, as a conceptual
object in order to think geometry, the south-pointing chariot presents a great interest.
So, we can ask what becomes this fundamental property of the south-pointing chariot when it moves on a sphere or
more generally on any arbitrary surface. The answer to this good question was proposed by Santander
\cite{sant} in 1992 in the limit of zero size of the chariot. He found that the pointer is parallel
transported along the curve of the surface describing the paths of the points of contact of the wheels.
A proof of this property has also been proposed in the appendix of the book of Foster et Nightingale
\cite{fost} in 1995.

Why we return to this problem? In the mentioned proofs, the paths of the wheels are identified to two
infinitely near curves considered as parallel. The study was done in the tangent plane to the
surface at each point along the curves. We think that it is desirable to take up this problem
by making a proof which takes into account the geometry of two-wheeled chariot and the fact that
both wheels have to roll without slipping on the surface. 

We firstly establish the general equations governing a two-wheeled chariot with both wheels
rolling without slipping. We solve them up to the first-order approximation in the second fundamental form 
of the surface.
The approximation of order zero leads the approximate solution of Santander \cite{sant}. On the contrary,
only some particular motions of the two-wheeled chariot are possible within the first-order approximation.

We secondly turn to the general equations governing this problem for the sphere. We show that a two-wheeled
chariot can always move with both wheels rolling without slipping. Then, we define a pointer for the south-pointing
chariot on the sphere. We prove that the pointer is parallel transported exactly. 

This paper is organized as follows. In section 2, we recall some elements of differential geometry of a surface.
We give in section 3 the general equations describing the motion of a two-wheeled chariot on a surface. We
specialize them to the case of the sphere in section 4. We develop in section 5 a procedure of successive
approximations. The south-pointing chariot is studied in section 6, in particular in the limit of zero 
size of the chariot. We return in section 7 to the case of the sphere for
determining exactly the properties of the south-pointing chariot. We add
some concluding remarks in section 8.

\section{Elements of differential geometry of a surface}

We recall some elements of differential geometry. We define a surface $(S)$ of the Euclidean space by
the following equation:
\begin{equation}\label{surf}
x^3=f(x^1,x^2)
\end{equation}
where the  $(x^i)$, i=1,2,3, are the Cartesian coordinates for an origin $O$. We adopt for the surface $(S)$ the
coordinate system $(x^A)$, $A=1,2$. The basis vectors $(\partial_A)$ of the tangent plane to the
surface $(S)$ at the point $(x^A)$ have the Cartesian components
$$
\vec{\partial}_1=\left( 1,0,\partial_1f\right) \quad {\rm and}\quad 
\vec{\partial}_2=\left( 0,1,\partial_2f\right) .
$$
We denote by $\vec{n}$ the unit normal vector to the surface $(S)$ and its components are
\begin{equation}\label{nor}
\vec{n}=\frac{1}{\sqrt{g}} \left( -\partial_1f,-\partial_2f,1\right) \quad {\rm with} \quad
g=1+\left( \partial_1f\right)^2+\left( \partial_2f\right)^2.
\end{equation}
In the coordinates $(x^A)$, the surface $(S)$ is provided with the induced metric $g_{AB}$ having for components
\begin{equation}\label{gab}
g_{AB}=\delta_{AB}+\partial_Af\partial_Bf ,
\end{equation}
where $\delta_{AB}$ is the Kronecker symbol,
and with a second fundamental form $k_{AB}$ of components
\begin{equation}\label{kab}
k_{AB}=\frac{1}{\sqrt{g}}\partial_{AB}f 
\end{equation}
where $g$ is the determinant of the metric $g_{AB}$.

The two principal curvatures $\sigma_1$ and $\sigma_2$ of the surface $(S)$ are defined by the following
inequalities:
$$
\sigma_1\leq \frac{k_{AB}X^AX^B}{g_{AB}X^AX^B} \leq \sigma_2
$$
for all vectors $X^A$. Indeed, $\sigma_1$ and $\sigma_2$ are the eigenvalues of the second fundamental
form $k_{AB}$ relative to the metric $g_{AB}$. 

We now define a curve $(\gamma )$ on the surface $(S)$. We give the parametric equation $x^A(s)$ where $s$
is the arc length. The unit tangent vector to the curve $(\gamma )$ is denoted $t^A$; we have
$t^A=dx^A/ds$. We point out that the curve $(\gamma )$ can be interpreted as a curve in the Euclidean space
by using the parametric equation $\vec{x}(s)$ where $x^3(s)=f(x^1(s),x^2(s))$, the unit vector tangent 
being $\vec{t}$.

We introduce the orthonormal frame of Darboux-Ribaucour $(\vec{t}, \vec{g}, \vec{n})$ defined along the curve
$(\gamma )$, $\vec{g}$ being defined by $\vec{g}=\vec{n}\wedge \vec{t}$. We can express the vectors of the
Euclidean space in this frame, in particular the derivative of the vectors of the
frame with respect to $s$. We have the well-known formulas
\begin{eqnarray}\label{darboux}
& & \frac{d\vec{t}}{ds}=\kappa_g\vec{g}+\kappa_n\vec{n} ,  \nonumber \\
& & \frac{d\vec{g}}{ds}=-\kappa_g\vec{t}+\theta_g\vec{n} , \\
& & \frac{d\vec{n}}{ds}=-\kappa_n\vec{t}-\theta_g\vec{g} , \nonumber
\end{eqnarray}
where $\kappa_g$ is called the geodesic curvature, $\kappa_n$ the normal curvature and $\theta_g$ the geodesic
torsion. By making use of components (\ref{nor}) of $\vec{n}$, we find after some calculations
\begin{equation}\label{kett}
\kappa_n=k_{AB}t^At^B \quad {\rm and} \quad \theta_g=k_{AB}g^At^B .
\end{equation}
Thus, $\kappa_n$ and $\theta_g$ depend on the second fundamental form. On the other hand, $\kappa_g$ depends
only of the curve $(\gamma )$ and of the metric $g_{AB}$. We have the identity
$$
g_{AB}t^A\frac{\nabla g^B}{ds}=\vec{t}\cdot \frac{d\vec{g}}{ds} ,
$$
where $\nabla$ is the covariante derivative associated with the metric $g_{AB}$. In consequence, we have
the formulas
$$
\frac{\nabla t^A}{ds}=\kappa_gg^A \quad {\rm and} \quad \frac{\nabla g^A}{ds}=-\kappa_gt^A .
$$
So, the curve $(\gamma_R)$ is a geodesic curve if and only if $\kappa_g=0$.

\section{Two-wheeled chariot on a surface}

\subsection{Rolling without slipping of a wheel}

We begin by studying a single wheel rolling on the surface $(S)$ defined by equation (\ref{surf}). 
We will examine the conditions such that this
wheel is rolling without slipping. We assume that the wheel is rigid and infinitely thin. It is
represented by a rotating disc around an axle characterized by an unit vector $\vec{\mu}$ perpendicular to it.
The point of contact of the disc with the surface $(S)$ is denoted by $Q$. Calling $P$ the center of the disc,
we set $\overrightarrow{QP}=R\vec{\tau}$ where $\vec{\tau}$ is an unit vector and of course we have
\begin{equation}\label{taumu}
\vec{\tau}\cdot \vec{\mu}=0 .
\end{equation}
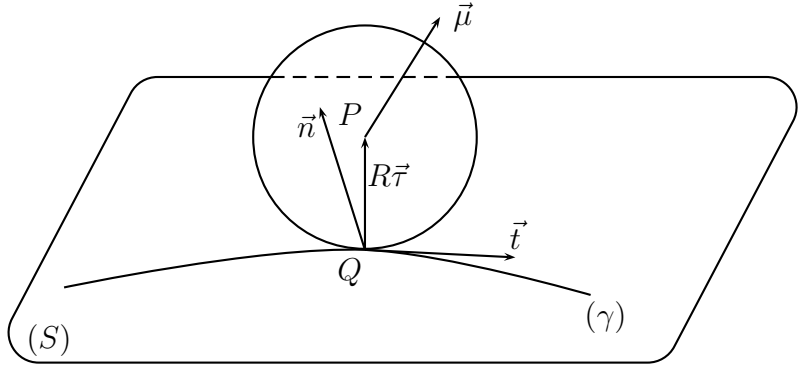
\begin{figure}[htb]
\hspace{2cm}
\begin{pspicture}(12,6)
\pscircle(5,4){1.5}
\pscurve(1,2)(5,2.5)(8,1.9)
\psline{->}(5,2.5)(7,2.4) \psline{->}(5,2.5)(5,4) \psline{->}(5,4)(6,5.6) \psline{->}(5,2.5)(4.4,4.4)
\rput(4.8,2.2){$Q$} \rput(4.8,4.3){$P$} \rput(0.8,1.3){$(S)$} \rput(5.3,3.5){$R\vec{\tau}$}
\rput(7,2.7){$\vec{t}$} \rput(8.2,1.6){$(\gamma )$} \rput(6.3,5.6){$\vec{\mu}$} \put(4.1,4){$\vec{n}$}
\psline[linearc=0.4](6.25,4.8)(11,4.8)(9,1)(0,1)(2,4.8)(3.75,4.8)
\psline[linestyle=dashed](3.75,4.8)(6.25,4.8)
\end{pspicture}
\caption{Rolling without slipping of a wheel on a surface $(S)$}
\end{figure}
We emphasize that the wheel is not necessarily perpendicular to the tangent plane at the point $Q$.
We recognize that we should take a wheel with a certain thickness in a realistic case but this is too difficult.

The points of contact of the rotating wheel define a curve $(\gamma )$ on the surface $(S)$. We take its equation
in the form $x^A(s)$, parametrized by the arc length $s$, with the unit tangent vector $t^A$. In Euclidean space,
this corresponds to $\vec{x}(s)$ and $\vec{t}$ with $\vec{t}\cdot \vec{n}=0$.

We require that the wheel rolls without slipping and without lateral skidding.
This means in particular that $\vec{t}$ is in the plane of the wheel and we have thereby
\begin{equation}\label{tmu}
\vec{t}\cdot \vec{\tau}=0 \quad {\rm and} \quad \vec{t}\cdot \vec{\mu}=0 ,
\end{equation}
and moreover the angle $\delta \theta$ of the rotating motion of the wheel, corresponding to the covered
distance $\delta s$ on the curve $(\gamma )$, is given by
$$
\delta \theta =\frac{1}{R}\delta s .
$$

\subsection{Two wheels on a common axle}

We now consider a chariot, with two wheels rotating independently on a common axle, which moves on
a surface $(S)$. The different quantities for the right wheel and the left wheel are indexed by $R$ and $L$
respectively. Each wheel must satisfy conditions (\ref{taumu}) and (\ref{tmu}) 
with a common vector $\vec{\mu}$, ensuring the rolling without slipping. We suppose that the motion of the chariot
is driven by the right wheel. We give thus the parametric equation $\vec{x}_{R}(s_R)$
of the curve $(\gamma_R)$ with the unit tangent vector $\vec{t}_R$. We will seek whether it is possible
to determine the point $Q_L$ of the surface $(S)$ and its path. We will establish the exact relations 
taking into account the geometry of the chariot and the fact that both wheels roll without slipping.

The two points $Q_R$ and $Q_L$ are directly connected by the geometry of the 
two-wheeled chariot. In vectorial notation, this relation can be written as
\begin{equation}\label{geom}
\overrightarrow{OQ_L}=\overrightarrow{OQ_R}+R\vec{\tau}_R+D\vec{\mu}-R\vec{\tau}_L .
\end{equation}
Since the points $Q_R$ and $Q_L$ are on the surface $(S)$, we have the following relations:
$x^{3}_{R}=f(x^{A}_{R})$ and $x^{3}_{L}=f(x^{A}_{L})$. 
We recall conditions $(\ref{taumu})$: 
$\vec{\tau}_R\cdot \vec{\mu}=0$ and $\vec{\tau}_L\cdot \vec{\mu}=0$. However, there is no reason
why the vectors $\vec{\tau}_R$ and $\vec{\tau}_L$ coincide. The geometry of the 
two-wheeled chariot and the conditions
of rolling without slipping determine in principle the point $Q_L$ which goes over the curve
$(\gamma_L)$. However, $s_R$ is not in general the arc length of the curve $(\gamma_L)$.
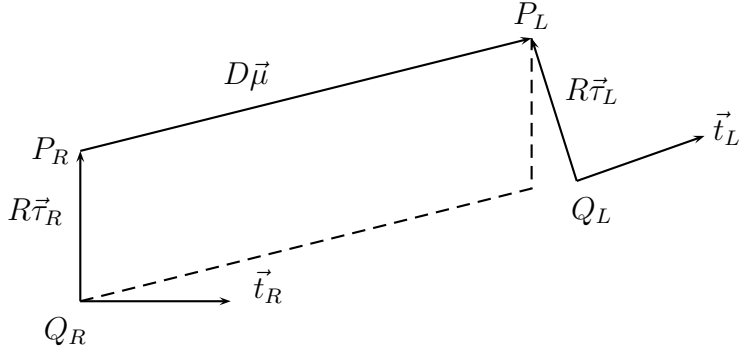
\begin{figure}[htb]
\hspace{2cm}
\begin{pspicture}(12,5)
\psline{->}(2,1)(2,3) \psline{->}(2,3)(8,4.5) \psline{<-}(8,4.5)(8.6,2.6)
\psline[linestyle=dashed](2,1)(8,2.5)(8,4.5)
\psline{->}(2,1)(4,1) \psline{->}(8.6,2.6)(10.3,3.2)
\rput(1.8,0.6){$Q_R$} \rput(4.5,1.2){$\vec{t}_R$} \rput(1.6,3){$P_R$}
\rput(8,4.8){$P_L$} \rput(8.8,2.2){$Q_L$} \rput(10.6,3.3){$\vec{t}_L$} 
\rput(4.2,4){$D\vec{\mu}$} \rput(1.4,2.2){$R\vec{\tau}_R$} \rput(8.8,3.8){$R\vec{\tau}_L$}
\end{pspicture}
\caption{Two-wheeled chariot on a surface $(S)$}
\end{figure}

We call $\vec{n}_R$ the unit normal vector to the surface $(S)$ along the curve $(\gamma_R)$.
We use the Darboux-Ribaucour frame $(\vec{t}_R, \vec{g}_R, \vec{n}_R)$ attached to the curve
$(\gamma_R)$. 
This one is characterized by $\left( \kappa_g\right)_R$, $\left( \kappa_n\right)_R$ and
$\left( \theta_g\right)_R$.
We first write down $\vec{\tau}_R$ with the help of an unknown parameter $\alpha$ by setting
\begin{equation}\label{taur}
\vec{\tau}_R=\cos \alpha \vec{n}_R+\sin \alpha \vec{g}_R
\end{equation}
which takes into account the fact that $\vec{\tau}_R$ is orthogonal to $\vec{t}_R$. From (\ref{taur}), we deduce
\begin{equation}\label{mu}
\vec{\mu}=-\sin \alpha \vec{n}_R+\cos \alpha \vec{g}_R
\end{equation}
since $\vec{\mu}$ is orthogonal to $\vec{\tau}_R$ and to $\vec{t}_R$. We secondly write down $\vec{\tau}_L$ 
with the help of an unknown parameter $\beta$ by setting
\begin{equation}\label{taul}
\vec{\tau}_L=\cos \beta \cos \alpha \vec{n}_R+\cos \beta \sin \alpha \vec{g}_R+\sin \beta \vec{t}_R
\end{equation}
since $\vec{\tau}_L$ is orthogonal to $\vec{\mu}$. By substituting
(\ref{taur}), (\ref{mu}) and (\ref{taul}) in expression (\ref{geom}) of
$\overrightarrow{OQ_L}$, we obtain the exact relation between the points
$Q_L$ and $Q_R$ in the Darboux-Ribaucour frame
\begin{eqnarray}\label{rela}
& & \overrightarrow{OQ_L}=\overrightarrow{OQ_R}+\left( -D\sin\alpha +R\cos \alpha -R\cos \beta \cos \alpha \right) \vec{n}_R  \nonumber \\
& & +\left( D\cos \alpha + R\sin \alpha -R\cos \beta \sin \alpha \right) \vec{g}_R -
R\sin \beta	\vec{t}_R .
\end{eqnarray}
For convenience, we define the vector $\vec{Y}$ by
\begin{equation}\label{y}
\overrightarrow{OQ_L}=\overrightarrow{OQ_R}+D\vec{Y} \quad {\rm or} \quad \overrightarrow{Q_RQ_L}=D\vec{Y} .
\end{equation}

The expression of $\vec{Y}$ depends on two unknown parameters $\alpha$ and $\beta$. We are going to determine them
by requiring on the hand that the point $Q_L$ belongs to the surface $(S)$, 
\begin{equation}\label{relay}
x^{3}_{R}+DY^3=f(x^{1}_{R}+DY^1,x^{2}_{R}+DY^2) ,
\end{equation}
and the other hand that the unit tangent vector $\vec{t}_L$ to the curve $(\gamma_L)$ should be orthogonal to
$\vec{\tau}_L$ and to $\vec{\mu}$. Differentiating expression (\ref{rela}) of
$\overrightarrow{OQ_L}$ with respect to $s_R$ yields the following tangent vector:
\begin{equation}\label{vit}
\vec{t}'_L=\vec{t}_R+D\frac{d\vec{Y}}{ds_R} .
\end{equation}
Conditions (\ref{tmu}) can be fulfilled on $\vec{t}'_L$ since the vector
$\vec{t}'_L$ is proportional to $\vec{t}_L$.

\section{Two-wheeled chariot on the sphere}

We now specialize the surface $(S)$ to be the sphere $S^2$. This one is centred around the origin $O$ and its
radius is $L$. We describe the hemisphere defined by $x^3\geq 0$ by the following function: 
\begin{equation}\label{fS2}
f(x^1,x^2)=\sqrt{L^2-\left( x^1\right)^2-\left( x^2\right)^2} .
\end{equation}
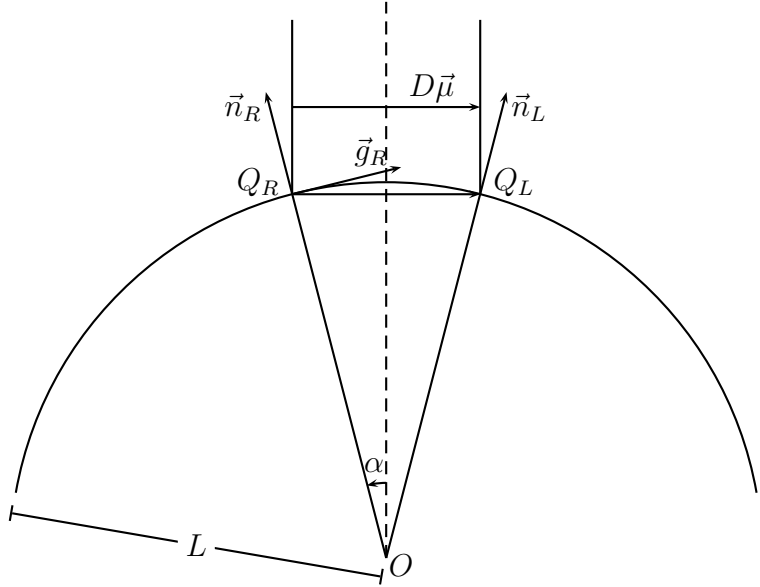
\begin{figure}[htb]
\hspace{2cm}
\begin{pspicture}(12,9)
\psarc(6,1){5}{10}{170}
\psline{->}(6,1)(4.4,7.2) \psline{->}(6,1)(7.6,7.2)
\psline{->}(4.75,5.84)(6.2,6.2) \psline{->}(4.75,7)(7.25,7) \psline{->}(4.75,5.84)(7.25,5.84)
\psline(4.75,5.84)(4.75,8.16) \psline(7.25,5.84)(7.25,8.16) 
\psline[linestyle=dashed](6,1)(6,8.4)
\rput(6.2,0.9){$O$} \rput(4.3,6){$Q_R$} \rput(7.7,6){$Q_L$} \rput(5.84,2.24){$\alpha$}
\rput(4.1,7){$\vec{n}_R$} \rput(7.9,7){$\vec{n}_L$} \rput(6.6,7.25){$D\vec{\mu}$} \rput(5.8,6.4){$\vec{g}_R$}
\pcline{|-|}(1,1.6)(5.95,0.75) \mput*{$L$}
\psarc{->}(6,1){1}{90}{105.1}
\end{pspicture}
\caption{Motion of a two-wheeled chariot on the sphere $S^2$}
\end{figure}
We derive immediately from (\ref{fS2}) the components $g_{AB}$ of the metric 
\begin{eqnarray}\label{gabS2}
& & g_{11}=\frac{L^2-\left( x^2\right)^2}{L^2-\left( x^1\right)^2-\left( x^2\right)^2} , \nonumber \\
& & g_{12}=\frac{x^1x^2}{L^2-\left( x^1\right)^2-\left( x^2\right)^2} , \\
& & g_{22}=\frac{L^2-\left( x^1\right)^2}{L^2-\left( x^1\right)^2-\left( x^2\right)^2} , \nonumber
\end{eqnarray}
and the second fundamental form
\begin{equation}\label{kabS2}
k_{AB}=-\frac{1}{L}g_{AB} .
\end{equation}
According to (\ref{kabS2}), we see that the two principal curvatures are equal, $\sigma_1=\sigma_2=-1/L$. We deduce
also that 
\begin{equation}\label{kn}
\kappa_n=-\frac{1}{L} \quad {\rm and}\quad \theta_g=0 
\end{equation}
for all the curves on the sphere $S^2$.

We take up exact relation (\ref{rela}) for a two-wheeled chariot. For reasons of symmetry, we have
$\vec{\tau}_L=\vec{\tau}_R$ and thus $\beta =0$. The angle $\alpha$, which characterizes expression (\ref{mu}) of
$\vec{\mu}$, is constant and we find by the geometry of the problem
\begin{equation}\label{alphaS2}
\sin \alpha =\frac{D}{2L} .
\end{equation}
Expression (\ref{rela}) reduces then to
$$
\overrightarrow{OQ}_L=\overrightarrow{OQ}_R-D\sin \alpha \vec{n}_R+D\cos \alpha \vec{g}_R .
$$

By using the Darboux-Ribaucour formulas (\ref{darboux}) for a sphere, {\em i.e.}  with (\ref{kn}), 
expression (\ref{vit}) of the tangent vector becomes
\begin{equation}\label{tlS2}
\vec{t}'_L=\left( 1-D\left( \kappa_g\right)_R \cos \alpha -\frac{D\sin \alpha}{L}\right) \vec{t}_R
\end{equation}
and, except in the case where $\vec{t}'_L=0$, we have
$$
\vec{t}_L=\vec{t}_R .
$$
We verify easily conditions (\ref{tmu}) for the left wheel.
So, a two-wheeled chariot can arbitrarly move with both wheels rolling without slipping on the sphere $S^2$.

\section{Approximate motion of a two-wheeled chariot}

Except in the previous case of the sphere, it is necessary to consider a procedure of 
successive approximations
for studying the motion of a two-wheeled chariot on any arbitrary  surface $(S)$. 
We consider the Taylor series of the function $f$ that we write up to the second order in a sufficiently small
neighbourhood of the point $x^{A}_{R}$. Equation (\ref{surf}) becomes
\begin{equation}\label{surfa}
x^{3}=f+\partial_{A}f\left( x^{A}-x^{A}_{R}\right) +\frac{1}{2}\partial_{AB}f
\left( x^{A}-x^{A}_{R}\right) \left( x^{B}-x^{B}_{R}\right) +\cdots
\end{equation}
where the coefficients $f$, $\partial_{A}f$ et $\partial_{AB}f$ are taken at the point $x^{A}_{R}$. Henceforth, 
equation (\ref{surfa}) is considered as the equation of the surface $(S)$ in the neighbourhood of the
point $x^{i}_{R}$. 
With the second fundamental form $k_{AB}$ of the surface $(S)$, we can rewrite (\ref{surfa}) in the form
\begin{equation}\label{surfk}
x^{3}=f+\partial_{A}f\left( x^{A}-x^{A}_{R}\right) +\frac{1}{2}\sqrt{g}k_{AB}
\left( x^{A}-x^{A}_{R}\right) \left( x^{B}-x^{B}_{R}\right) +\cdots .
\end{equation}

We have two groups of conditions for determining $\vec{Y}$ by successive approximations.
\begin{enumerate}
\item The point $Q_L$ belongs to $(S)$. Thus, we must verified  relation (\ref{relay}) within our approximation.
We rewrite (\ref{surfk}) at the point $Q_L$ in the following form:
\begin{equation}\label{relaa}
\vec{n}_R\cdot \vec{Y}=\frac{D}{2}k_{AB}Y^AY^B +\cdots
\end{equation}
in which we have introduced the unit normal vector $\vec{n}_R$ with components (\ref{nor}).
\item The two conditions (\ref{tmu}) for the left wheel.
\end{enumerate}

In our procedure of successive approximations, the right member of equation (\ref{relaa}) is the first
perturbation. Its value verifies the inequalities
$$
D\sigma_1 \leq \frac{Dk_{AB}Y^AY^B}{g_{AB}Y^AY^B} \leq D\sigma_2 .
$$
We similarly expand $\vec{Y}$ under the form
\begin{equation}\label{y01}
\vec{Y}=\vec{Y}_0+\vec{Y}_1+\cdots 
\end{equation}
where the term $\vec{Y}_1$ is the first perturbation of order $Dk_{AB}$.

\subsection{Zeroth-order approximation}

If we can strictly neglect $D$ in comparison with the absolute values of $1/\sigma_1$ and $1/\sigma_2$, then 
the insertion of form (\ref{y01}) into equation (\ref{relaa}) yields
$$
\vec{n}_R\cdot \vec{Y}_0=0 .
$$
In this approximation, equation (\ref{surfa}) of the surface $(S)$ reduces to the one of a plane 
$$
x^3=f+\partial_Af\left( x^A-x_{R}^{A}\right) .
$$
Consequently, we start with the solution of our problem valid in the plane which is characterized by
\begin{eqnarray}\label{y0}
& & \vec{Y}_0=\vec{g}_R ,\nonumber \\
& & \vec{\mu}=\vec{g}_R , \\
& & \vec{\tau}_R=\vec{\tau}_L=\vec{n}_R .\nonumber
\end{eqnarray}
Since $D\theta_g$ and $D\kappa_n$ are neglected, the tangent vector $\vec{t}'_L$ given by (\ref{vit}) reduces to
\begin{equation}\label{tl0}
\vec{t}'_L=\left( 1-D\left( \kappa_g\right)_R \right) \vec{t}_R .
\end{equation}
Conditions (\ref{tmu}) with vector (\ref{tl0}) are obviously verified. In the case $\left( \kappa_g\right)_R=1/D$,
the left wheel revolves around the point $Q_L$.

The motion of the two-wheeled chariot with both wheels rolling without slipping is always possible in the 
zeroth-order approximation. The tangent planes along the curves $(\gamma_R)$ and $(\gamma_L)$ are
identical at this approximation. The two wheels are always perpendicular to this tangent plane.
We can calculate the geodesic curvature of the curve $(\gamma_L)$,
$$
\left( \kappa_g\right)_L=\frac{\left( \kappa_g\right)_R}{1-D\left( \kappa_g\right)_R} .
$$

\subsection{First-order approximation}

We retain only the terms of first order in $Dk_{AB}$ in the general equations of the problem. By virtue of
solution (\ref{y0}) at the order zero,
we expand expression (\ref{rela}) of $\vec{Y}$ by assuming that the parameters $\alpha$ and $\beta$ are terms of
order $Dk_{AB}$, denoted $\alpha_1$ and $\beta_1$. According to (\ref{y01}), we write down
\begin{equation}\label{ya}
\vec{Y}=\vec{g}_R+\vec{Y}_1 \quad {\rm with} \quad \vec{Y}_1=-\alpha_1\vec{n}_R-\frac{R}{D}\beta_1\vec{t}_R ,
\end{equation}
and also the other vectors
\begin{equation}\label{mua}
\vec{\mu}=\vec{g}_R -\alpha_1\vec{n}_R ,
\end{equation}
\begin{equation}\label{taua}
\vec{\tau}_L=\vec{n}_R +\alpha_1\vec{g}_R+\beta_1\vec{t}_R .
\end{equation}

The first condition is the requirement that the point $Q_L$ belongs to the surface $(S)$. According to 
(\ref{relaa}), we get an  equation governing $\vec{Y}_1$ 
\begin{equation}\label{relaa1}
\vec{n}_R\cdot \vec{Y}_1=\frac{D}{2}k_{AB}g^{A}_{R}g^{B}_{R} .
\end{equation}
By substituting (\ref{ya}) in (\ref{relaa1}), we find immediately
\begin{equation}\label{alpha}
\alpha_1=-\frac{D}{2}k_{AB}g^{A}_{R}g^{B}_{R}  
\end{equation}
and so the parameter $\alpha_1$ is determined.

The second group of conditions needs the calculation of $\vec{t}'_L$. Expression (\ref{vit}) 
with $\vec{Y}$ given by (\ref{ya}) reduces to
\begin{equation}\label{vita}
\vec{t}'_L=\vec{t}_R-D\left( \kappa_g\right)_R\vec{t}_R+D\left( \theta_g\right)_R
\vec{n}_R-D\frac{d\alpha_1}{ds_R}\vec{n}_R-R\frac{d\beta_1}{ds_R}\vec{t}_R-
R\left( \kappa_g\right)_R\beta_1\vec{g}_R ,
\end{equation}
retaining only the terms of order $Dk_{AB}$. Within the same approximation, 
we calculate the derivative of $\alpha_1$ with respect to $s_R$
\begin{equation}\label{dalpha}
\frac{d\alpha_1}{ds_R}=D\left( \kappa_g\right)_Rk_{AB}t^{A}_{R}g^{B}_{R}  
-\frac{D}{2}t^{C}_{R}\partial_Ck_{AB}g^{A}_{R}g^{B}_{R} .
\end{equation}
However, when we insert (\ref{dalpha}) into (\ref{vita}), we can
neglect the terms of order in $D^2\partial_Ck_{AB}$ in comparison with the ones of order in $Dk_{AB}$. 
Finally, we obtain
\begin{equation}\label{vitabis}
\vec{t}'_L=\left( 1-D\left( \kappa_g\right)_R -R\frac{d\beta_1}{ds_R}\right) \vec{t}_R+
\left( 1-D\left( \kappa_g\right)_R\right) D\left( \theta_g \right)_R\vec{n}_R
-R\left( \kappa_g\right)_R\beta_1\vec{g}_R .
\end{equation}

We are now in a position to verify the two conditions (\ref{tmu}) for the left wheel.
\begin{enumerate}
\item The scalar product $\vec{t}'_L\cdot \vec{\tau}_L=0$.
By using (\ref{taua}) and (\ref{vitabis}), we then obtain the condition
\begin{equation}\label{eqb2}
\left( 1-D\left( \kappa_g\right)_R\right) \beta_1 +\left( 1-D\left( \kappa_g\right)_R\right) 
D\left( \theta_g\right)_R=0
\end{equation}
and so the parameter $\beta_1$ is determined
\begin{equation}\label{beta}
\beta_1=-D\left( \theta_g\right)_R.
\end{equation}
\item The scalar product $\vec{t}'_L \cdot \vec{\mu}=0$.
By using (\ref{mua}) and (\ref{vitabis}), we obtain
\begin{equation}\label{tlmu}
\vec{t}'_L \cdot \vec{\mu}=-\beta_1R\left( \kappa_g\right)_R .
\end{equation}
\end{enumerate}

Expression (\ref{tlmu}) vanishes either $\beta_1=0$, equivalent to $\left( \theta_g\right)_R=0$, or 
$\left( \kappa_g\right)_R=0$.
The crucial consequence is that a two-wheeled chariot moves, with both wheels rolling without slipping, 
only on particular paths: either a line of curvature or a geodesic curve. This severe restriction does not exist
in the zeroth-order approximation. It does not apply to the sphere because all the curves satisfy $\theta_g=0$.

\section{South-pointing chariot on a surface}

We are now going to consider a south-pointing chariot on a surface $(S)$, equipped with a differential gear
as explained in the introduction. The first thing to be assumed is that both wheels roll without slipping.
Otherwise, we will lose the relation between the angle of the rotating motion of the wheels and the
covered distance on the surface $(S)$. We use the results of the previous section.

\subsection{Zeroth-order approximation}

This approximation corresponds to the case where we can strictly neglect $D$ in comparison with the absolute values
of $1/\sigma_1$ and $1/\sigma_2$ of the surface $(S)$.
Having determined the tangent vector (\ref{tl0}), we can express $ds_L$ in function of $ds_R$ 
\begin{equation}\label{slsr0}
ds_L=\left( 1-D\left( \kappa_g\right)_R\right) ds_R .
\end{equation}
To avoid unessential complications, we assume that $D\left( \kappa_g\right)_R <1$.
The infinitesimal angles of the rotating motion of the right and left wheel are respectively
$$
d\theta_R=\frac{1}{R}ds_R \quad {\rm and} \quad d\theta_L=\frac{1}{R}ds_L .
$$
According to (\ref{diff1}), the differential gear gives continually an angle $d\phi$ given by
\begin{equation}\label{dphi0}
d\phi =\frac{D}{2R}\left( \kappa_g\right)_Rds_R .
\end{equation}
As in the plane, we choose $D=2R$ to obtain the differential expression
\begin{equation}\label{dphids}
\frac{d\phi}{ds_R}=\left( \kappa_g\right)_R .
\end{equation}

We define the pointer as a vector $p^A$ of the tangent plane to the surface $(S)$ along the curve $(\gamma_R)$
which makes an angle $\phi$ with the axle of the wheels. We thus set
\begin{equation}\label{pa0}
p^A=\sin \phi t^{A}_{R}+\cos \phi g^{A}_{R}
\end{equation}
and the variation of the angle $\phi$ is given by (\ref{dphids}). The pointer
is located in the plane containing the axle of the wheels so that this plane remains orthogonal to 
normal vector $\vec{n}_R$ along the curve $(\gamma_R)$.

By the differential geometry of the surface $(S)$, we have the identity
$$
\frac{\nabla p^A}{ds_R}=\left( \frac{d\phi}{ds_R}-\left( \kappa_g\right)_R\right) \left( \cos \phi t^{A}_{R}
-\sin \phi g^{A}_{R}\right) .
$$
Thus, we see from result (\ref{dphids}) that
\begin{equation}\label{para}
\frac{\nabla p^A}{ds_R}=0 .
\end{equation}
In the zeroth-order approximation, the pointer is parallel transported along $(\gamma_R)$, 
or equivalently along the curve $(\gamma_L)$. This is the result of Santander \cite{sant}.

\subsection{First-order approximation}

In this approximation, we keep the terms of order in the absolute values of $D\sigma_1$ and $D\sigma_2$
of the surface $(S)$ but these terms are small in comparison with 1. From the results of subsection 5.2, 
the two wheels of the south-pointing chariot
roll without slipping if the curve $(\gamma_R)$ is such that 
$\left( \kappa_g\right)_R=0$ or $\left( \theta_g\right)_R=0$. The two wheels are not in the same tangent
plane. Also, there exists a crucial difficulty to define a pointer in the tangent plane along a certain curve
to be determined. We do not know it generally.

Fortunately, we can define such a pointer in the case of the sphere $S^2$ where all the curves satisfy 
$\theta_g=0$. We do not carry out the calculation within the first-order approximation since we give the exact
determination in the next section.

\section{South-pointing chariot on the sphere}

We have seen in section 4 that a two-wheeled chariot can arbitrarly moves with both wheels rolling
without slipping on the sphere $S^2$. We now characterize its motion by the curve $(\gamma_M)$
describing the path of the middle of the arc of circle between the points $Q_R$ and $Q_L$, denoted $M$.
This point $M$ is the projection of the middle of the axle of the wheels on the sphere.

We easily find by the geometry of the problem
\begin{equation}\label{qm}
\overrightarrow{Q_RM}=\parallel \overrightarrow{Q_RM}\parallel \left( -\sin \frac{\alpha}{2}\vec{n}_R+
\cos \frac{\alpha}{2}\vec{g}_R \right)
\end{equation}
with the norm
\begin{equation}\label{nqm}
\parallel \overrightarrow{Q_RM}\parallel =2L\sin \frac{\alpha}{2} .
\end{equation}
The tangent vector $\vec{t}'_M$ to the curve $(\gamma_M)$ is obtained by performing the differentiation of 
(\ref{qm}) with respect to $s_R$. Since $\alpha$ is a constant, we get
$$
\vec{t}'_M=\vec{t}_R+\parallel \overrightarrow{Q_RM}\parallel \left( -\sin \frac{\alpha}{2}
\frac{d\vec{n}_R}{ds_R}+\cos \frac{\alpha}{2}\frac{d\vec{g}_R}{ds_R} \right) .
$$
The Darboux-Ribaucour formulas for the curve $(\gamma_R)$ are expressed 
by taking (\ref{kn}). With the formulas $\cos \alpha =1-2\sin^2 \alpha /2$ and
$\sin \alpha =2\sin \alpha /2 \cos \alpha /2$, we obtain thereby
\begin{equation}\label{tm}
\vec{t}'_M=\left( \cos \alpha -L\left( \kappa_g\right)_R \sin \alpha \right) \vec{t}_R .
\end{equation}
We denote $\vec{t}_M$ the unit tangent vector to the curve $(\gamma_M)$ parametrized by $s_M$.

We deduce from (\ref{tm}) the useful formula
\begin{equation}\label{srsm}
\frac{ds_R}{ds_M}=\frac{1}{\cos \alpha -L\left( \kappa_g\right)_R\sin \alpha} .
\end{equation}
We need the expression of the geodesic curvature $\left( \kappa_g\right)_M$ for the curve
$(\gamma_M)$. To do this, we introduce the Darboux-Ribaucour frame $(\vec{t}_M,\vec{g}_M,\vec{n}_M)$
attached to the curve $(\gamma_M)$. We have the expressions
\begin{eqnarray}\label{tgnm}
& & \vec{t}_M=\vec{t}_R , \nonumber \\
& & \vec{g}_M=\cos \alpha \vec{g}_R-\sin \alpha \vec{n}_R , \\
& & \vec{n}_M=\sin \alpha \vec{g}_R+\cos \alpha \vec{n}_R . \nonumber 
\end{eqnarray}
On the sphere, $\left( \kappa_g\right)_M$ is defined by the following differential equation:
$$
\frac{d\vec{g}_M}{ds_M}=-\left( \kappa_g\right)_M\vec{t}_M .
$$
Now, 
$$
\frac{d\vec{g}_M}{ds_M}=\left( \frac{ds_R}{ds_M}\right) \frac{d\vec{g}_M}{ds_R}=
\left( \frac{ds_R}{ds_M}\right) \left( -\left( \kappa_g\right)_R\cos \alpha \vec{t}_R-
\frac{\sin \alpha}{L}\vec{t}_R\right) ,
$$
therefore we find
\begin{equation}\label{kgm}
\left( \kappa_g\right)_M=\frac{\displaystyle \left( \kappa_g\right)_R\cos \alpha +\frac{\sin \alpha}{L}}
{\cos \alpha -L\left( \kappa_g\right)_R\sin \alpha} .
\end{equation}

We now consider a south-pointing chariot on the sphere $S^2$. The pointer will be a vector of the tangent 
plane to the sphere along the curve $(\gamma_M)$. We set
\begin{equation}\label{pAS2}
p^A=\sin \phi t^{A}_{M}+\cos \phi g^{A}_{M} .
\end{equation}
The pointer defined by (\ref{pAS2}) is well connected with the geometry of the south-pointing chariot since
$\vec{g}_M$ coincides with $\vec{\mu}$, the axle of the wheels. We must suppose that the plane in
which the pointer rotates is orthogonal to the normal vector $\vec{n}_M$ along the curve $(\gamma_M)$.

Since the south-pointing chariot moves with both wheels rolling without slipping, the differential gear 
gives continually the infinitesimal angle 
$$
d\phi =\frac{1}{2R}\left( ds_R-ds_L\right) .
$$
According to (\ref{tlS2}) in section 4, we obtain
$$
d\phi =\frac{D}{2R}\left( \left( \kappa_g\right)_R\cos \alpha +\frac{\sin \alpha}{L}\right) ds_R .
$$
However, we must express $d\phi$ in function of $ds_M$. We have the differential expression
\begin{equation}
\frac{d\phi}{ds_M}=\frac{D}{2R}\left( \frac{ds_R}{ds_M}\right) \left( \left( \kappa_g\right) \cos \alpha +
\frac{\sin \alpha}{L}\right) .
\end{equation}
Taking into account (\ref{srsm}) and (\ref{kgm}), we obtain finally
\begin{equation}\label{dphiS2}
\frac{d\phi}{ds_M}=\frac{D}{2R}\left( \kappa_g\right)_M .
\end{equation}
With the choice $D=2R$ as in the plane case, (\ref{dphiS2}) reduces to
\begin{equation}\label{final}
\frac{d\phi}{ds_M}=\left( \kappa_g\right)_M .
\end{equation}
Hence, the vector $p^A$ defined by (\ref{pAS2}) is parallel transported along the curve $(\gamma_M)$,
\begin{equation}\label{dpAS2}
\frac{\nabla p^A}{ds_M}=0 .
\end{equation}
We emphasize that (\ref{dpAS2}) is an exact result on the sphere $S^2$. 

\section{Conclusion}

Our investigation about the south-pointing chariot on a surface, taking into account the rolling without 
slipping of both wheels, presents some new features with respect to the work of Santander \cite{sant}.
In the limit where the distance between the wheels tends to zero, we have found again that the south-pointing
chariot can move and that the pointer is parallel transported. On the contrary, within the first-order
approximation where the second fundamental form of the surface appears, only some particular
motions are possible: lines of curvature or geodesic curves. 

An interesting case is the one of the sphere. We have determined exactly the motion of the south-pointing
chariot on the sphere. We have defined a pointer along a specific curve on the sphere and we have proved that the 
pointer is parallel transported. Thus, we have extended the result of Santander \cite{sant} in an
exact manner to the case of the sphere.


\begin{thebibliography}{99}

\bibitem{need} NEEDHAM, J {\em The South-Pointing Carriage}, in Science and Civilisation in China, Vol. 4,
Part. II: Mechanical Engineering, Cambridge University Press (1965) pp. 286-303.
\bibitem{jing} JINGYAN, L {\em Studies of the South-Pointing Chariot: Survey of the Past 80 Years}, in
Chinese Studies in History and Philosophy of Science and Technology, eds Dainian and Cohen, Kluwer Academic
Publisher (1996) pp. 267-278.
\bibitem{lanc} LANCHESTER, G {\em The Yellow Emperor's South-Pointing Chariot}, avec une note de MOULE, A C,
The China Society (1947) 8 pages.
\bibitem{yan} YAN, H S {\em South-pointing Chariots}, in Reconstruction Designs of Lost Ancient Chinese
Machinery, Springer (2007) pp. 199-268.
\bibitem{coal} COALES, J F {\em Historical and Scientific Background of Automation}, Engineering {\bf 182}
(1956) 363-370.
\bibitem{sant} SANTANDER, M {\em The Chinese South-Seeking chariot: A simple mechanical device for visualizing
curvature and parallel transport}, Am. J. Phys. {\bf 60} (1992) 782-787.
\bibitem{fost} FOSTER, J et NIGHTINGALE, J {\em The Chinese connection}, in A Short Course in
General Relativity, Springer (1995) pp. 213-219.

\end{thebibliography}
\end{document}